# Estimating causal effects of sanctions impacts: what role for country-level studies?

Francisco Rodríguez[1]

1. Introduction

Do economic sanctions negatively impact living conditions in target countries? Are their effects large enough to cause or exacerbate acute suffering among members of vulnerable groups? Do they help produce changes in the conduct of targeted entities, and if so, do their potential benefits in these cases outweigh the costs inflicted on target populations? Has the shift to targeted sanctions observed over the past decades attenuated the negative consequences of coercive measures on living standards?

These questions stand at the center of much of contemporary research on sanctions. An extensive literature that goes back at least to the seminal work has aimed to use quantitative econometric and calibration methods to assess the effects of sanctions on living standards in target countries as well as on the behavior of targeted governments and entities.[2] Nevertheless, any attempts to answer these questions through the use of quantitative statistical methods applied to cross-national or country-level data must confront a set of formidable challenges to empirical identification. Both the imposition of sanctions and the evolution of the target country's living standards are endogenous variables that form part of complex national and supra-national processes and which simultaneously impact and are impacted by a multiplicity of political, social, economic and cultural variables. Convincingly teasing out the links of causation in these relationships often requires data that satisfy conditions rarely present in non-experimental settings.

This article reviews recent advances in addressing these empirical identification issues in cross-country and country-level studies. I argue that, given the difficulties in assessing causal relationships in cross-national data, country-level case studies can serve as a useful and informative complement to cross-national regression studies. However, I also warn that case studies pose a set of additional potential empirical pitfalls which can obfuscate rather than clarify the identification of causal mechanisms at work. Therefore, the most sensible way to read case study evidence is as a complement rather than as a substitute to cross-national research.

2. Correlation and its discontents.

Every student who has been through an introductory statistics course will have heard the warning that correlation does not equal causation. With little hint of irony, the same student will then have spent the rest of the semester learning statistical models designed to estimate partial correlations and have gained little clue as to how to distinguish these from genuine causal effects. Even more advanced students who obtain some familiarity with econometric identification techniques will often only learn methods for identifying causal effects that work under stringent conditions hard to satisfy in many real-world applications.

---

[1] Francisco.rodriguez4@du.edu. Josef Korbel School of International Studies, University of Denver.
[2] Hufbauer et. al. (1990).

Classical statistical analysis is based on the idea that researchers know the variables in the model, the functional form of the relationship and the distribution of the errors prior to estimating a regression. Once quipped, "this advice could be treated as ludicrous, except that it fills all the econometric textbooks."[3] The inconvenient truth is that we don't know whether our coefficient estimates are biased, nor by how much, unless we know whether our model includes all relevant variables, omits all irrelevant ones, and has a correctly specified functional form. But this is precisely what we are trying to find out.

There is a fundamental difference between the experimental designs frequently used in natural sciences and the econometric analysis of non-experimental data ubiquitous to many social science settings. An adequately designed experiment can attempt to hold all other variables except the treatment of interest constant. Doing so is impossible in the non-experimental data that we have to address many questions of interest in the social sciences.

Leamer's critique is often credited with spurring the rise of what came to be known as the credibility revolution in empirical economics.[4] Key contributions led to the development of quasi-experimental methods that seek to replicate experimental research designs using non-experimental methods. A burgeoning literature has mainstreamed the use of designs such as difference-in-differences, matching, regression discontinuity, synthetic controls and machine learning. Another strand of the literature led to the design of randomized control trials (RCTs) to address causal identification issues in the social sciences.[5]

Notwithstanding its impact on many social science fields of study, the credibility revolution has yet to fully conquer cross-country econometric studies. In contrast to microeconomic studies, it is impossible to design randomized control trials to assess the effects of country-level interventions such as economic sanctions. We have yet to see a state sign off its sovereignty to accept being randomly sanctioned, nor, to the best of our knowledge, has any sanctions sender decided to randomize its decisions to freeze assets. Furthermore, the data requirements necessary to apply many quasi-experimental methods are sometimes so stringent as to preclude application in samples that are often limited to several dozen countries.

To illustrate the problems involved, imagine you are interested in studying whether sanctions negatively impact a target country's economic performance. Your first idea may be to run a regression in which the dependent variable is the target's GDP growth averaged over a certain period of time, and the explanatory variables include an indicator of whether the economy was sanctioned or not in that same period. Let's assume you also throw in a set of other controls that are known from the cross-national literature to be correlated with growth. An obvious problem with that regression comes from the fact that the countries that are most likely to end up being sanctioned could also be countries that would have seen weaker economic performance even in the absence of sanctions. The evidence that sanctions are correlated with growth could simply indicate that poorly performing economies are more likely to be those that have governments that do things likely to get them sanctioned.

---

[3] Leamer (1983, p. 36).
[4] Angrist and Pischke (2010); Deaton (2010).
[5] Some useful recent surveys of these methods include Cunningham (2021), Huntington-Klein (2021) and Angrist and Pischke(2009).

Suppose now that in response to this problem you decide to look at whether *changes* in growth rates are correlated with the imposition of sanctions. Under some conditions this may solve your problem; yet under others it could make it worse. For example, it may be that the countries that are more likely to be sanctioned are also those whose economic policies cause economic growth to deteriorate over time, for example, because the negative productivity effects of those policies increase the longer the policies remain in place.

Imagine now that, to address these concerns, you decide to look at whether changes in target countries' growth preceded the imposition of sanctions. Satisfied that you have nailed a causal explanation, you decide to proudly present your results at a professional conference. At the end of your talk, a macroeconomist raises her hand and observes that many of the countries that you are looking at suffered collapses in their stock markets just as investors became increasingly concerned that sanctions would be imposed. The fact that growth fell before sanctions were imposed simply shows that markets were able to anticipate what was coming.

Any attempts to infer causal relationships from non-experimental data must contend with two types of biases. One is that both your dependent and independent variable may be affected by a third variable that explains both. This is the case, for example, when poor policies explain both sanctions and economic performance. Economists refer to this phenomenon as confounder, or omitted variable, bias, because excluding the variable from a regression causes us to attribute a causal effect when there is none. Alternatively, both the dependent and independent variable may impact a third factor; for example, both sanctions and low growth may lead to capital flight. In this case, controlling for capital flight may introduce a spurious correlation between sanctions and growth even though no such relationship exists. This is known as collider bias, in which including a variable that should be excluded from our regression distorts coefficient estimates by absorbing part of the true causal effect that we are trying to estimate.

Attempting to infer causal relationships from cross-national data is an exercise fraught with difficulties. In other areas of economics, cognizance of these difficulties has sparked a preference for the use of randomized control trials that create genuine experimental settings in which confounders can be held invariant and colliders can be safely ignored. In non-experimental settings, they have led to the use of quasi-experimental methods that attempt to replicate experimental settings by comparing outcomes in exposed units with those of a non-exposed comparison group that can closely replicate the control group of an experiment.

Some recent studies apply quasi-experimental methods to analyze the effect of sanctions on economic growth and poverty. Comparing economies that were sanctioned by the United Nations Security Council (UNSC) with countries that were "nearly-sanctioned", that is, where a security council resolution to impose sanctions failed due to the veto of at least one permanent UNSC member.[6] The basic premise of this strategy is that nearly-sanctioned countries can be assumed to be similar to sanctioned countries and thus serve as an adequate control group. The authors find that while UN sanctions that were actually imposed had a negative and significant effect on economic growth, sanctions that failed to be imposed due to a veto have no meaningful correlation with growth. consider the effect of US economic sanctions on poverty by using matching methods to construct a counterfactual control group of countries that are as similar as possible to the sanctioned countries in a set of pre-sanctions characteristics.[7] They find that the poverty gap is 3.8 percentage points of GDP higher when a country is sanctioned by the United States. Comparing sanctioned countries with countries that were threatened with sanctions but where these sanctions were never imposed.[8] They find that the imposition of sanctions leads to a reduction of GDP growth of around 2.8 percentage points in the first two years after the imposition of sanctions, with no indication of a recovery even after sanctions are lifted.

---

[6] Neuenkirch and Neumeier (2015).

No application of quasi-experimental methods is perfect, as it is always possible to question the plausibility of the premises underlying their estimation. Countries that were able to enlist the help of a powerful ally willing to use its veto power to block a UNSC resolution may be systematically different from those that had no such friends, and it may be the ability and willingness of those allies to continue their economic ties with the near-sanctioned countries that explains their economic resilience. Similarly, the fact that some countries were threatened with sanctions, but never eventually sanctioned may reflect that their leaders altered their behavior because of their threats, and it may be that change in conduct that ultimately explains why their economies did better than those countries that ended up being sanctioned.

Nevertheless, the results of this strand of research are encouraging because they point in the same direction as that of many papers that do not use these quasi-experimental methods and have also found negative effects of sanctions on living standards. They thus add to what, taken together, starts to form a sizable array of evidence consistent with the hypothesis that sanctions have significant adverse negative effects on living conditions in target countries. Summarizing the results of 31 papers that have studied the effect of sanctions on indicators of living standards using quantitative statistical or model calibration methods.[9] Of these, 30 studies found unambiguous negative impacts on the variable of interest, one study found ambiguous (i.e., some positive and some negative) effects and only one study found unambiguous positive effects.[10]

3. **The economics of reasonable doubt**

---

[7] Neuenkirch and Neumeier (2016).
[8] Gutmann, et. al. (2021).
[9] Rodríguez (2023).
[10] The paper that finds ambiguous effects is Gutmann et al. (2018), which finds that US sanctions are associated with a deterioration of political rights but an improvement in women's emancipatory rights. The paper that finds unambiguous positive effects is Equipo Anova (2021), which argues that US sanctions on Venezuela were associated with improvements in imports of food and medicines. In Rodríguez (2022a) I show that the Equipo Anova (2021) results are driven by the omission of cereals and oils from their food imports variable and the questionable use of a linear specification that implies a counterfactual of negative imports in the absence of sanctions.

An old joke in the profession tells of an economist looking for his keys at night under a lamppost. When someone asks him where he lost the keys, the economist responds that he dropped them across the street. Why then is he looking for them under the lamppost, the bystander asks? "Because this is where the light is," the economist answers.

While the anecdote has been used by economists for decades, it fits quite well as a description of the current state of development economics. A 2015 analysis found that 38 percent of development papers published in the top eight economics journals were randomized controlled trial experiments.[11] Roughly over the same period, cross-country studies of the determinants of economic growth fell rapidly out of style, with even some of their most prominent users producing scathing critiques.[12] While, as the examples cited in the previous section show, it is possible to apply quasi-experimental methods to cross-country analysis, it is relatively rare for such studies to figure prominently among the most prominent examples of state-of-the art research in the profession. Of the 31 papers surveyed, none of them were published in one of the top eight economics journals, and only one of them appeared in a leading field journal.[13]

One of the reasons why cross-national studies have proved to be much less fertile ground for the use of quasi-experimental methods is their limited number of observations. While non-experimental studies in labor or health economics are typically conducted with observations of tens of thousands of individuals, allowing for credible estimation of individual, time, and cohort effects, the typical cross-country study will with luck have data on around 100 economies.

While none of the studies surveyed use the gold standard of RCT methodology, and only a handful use quasi-experimental methods, they nevertheless constitute the best evidence currently available to us on the human consequences of economic sanctions.[14] Thus, while it is certainly possible to construct alternative hypotheses to explain some of these data patterns and shed reasonable doubt on whether any individual estimation exercise has captured the causal relationships with the confidence that an RCT experiment could have done, it is also true that as of this time, the preponderance of the evidence is strongly consistent with the thesis that sanctions cause significant deteriorations in the living conditions of vulnerable groups in target economies. Given the magnitude of the consequences for the populations potentially affected by sanctions, these results suggest that policymakers need to at the very least exercise extreme caution in the use of this instrument and devote substantial time and resources to understanding how its potential negative effects can be mitigated.

4. **What role for case studies?**

Precisely because of the methodological and data limitations of cross-country regression studies, country-level case studies may have a particularly important role to play in this field.

---

[11] McKenzie (2016).
[12] Durlauf (2009) and Rodrik (2012).
[13] Rodríguez (2023) and Neuenkirch and Neumeier (2016).
[14] Rodríguez (2023).

Importantly, case studies allow us to study in detail the channels of causation through which the effects of sanctions may propagate through a specific economy or set of economies. Since these channels may be contingent on an economy's structural characteristics, there may be little hope of being able to capture them adequately in cross-country regression exercises. However, an adequately constructed case study may help us more closely and consistently trace the effect of these causes throughout the economy.

Consider, for example, the evolution of oil revenues and GDP shown in Figure 1 for three oil-exporting economies impacted during periods that include the imposition and – in some cases – lifting of sanctions: Iraq (1981-2005), Iran(2008-present) and Venezuela (2008-present). Data on each of these countries is insufficient to carry out a full-fledged regression analysis that adequately addresses the specification concerns highlighted in the previous sections. Yet visual inspection of these figures suggests that there is a strong case to be made for the hypothesis that sanctions negatively affected economic growth via lower export revenues. If we complement this finding with the knowledge that in all three countries oil exports accounted for the bulk of the country's export and fiscal revenues prior to sanctions, and that revenue shortfalls correlate strongly with decreases in public health and infrastructure spending, we see the contours of a story emerge in which the severing of links with the global economy profoundly affected living conditions in these target countries.

Case studies should not be seen as substitutes for cross-country regression work. They need to be treated as complements, with both types of studies forming part of the body of evidence used to assess the effect that sanctions can have on living conditions. While the evidence arising from a case study can be consistent with specific hypotheses about the effects of sanctions, it will typically also be consistent with multiple other hypotheses. Our judgment on whether to lend plausibility to those alternative hypotheses should be informed by knowledge on how frequently we see support for the same hypotheses emerge in comparable settings drawn from other case studies or cross-country empirical research.

5. **Revisiting the evidence on Venezuela sanctions**

In August 2017, the Trump administration imposed financial sanctions barring Venezuela or its state-owned companies from issuing or refinancing debt and receiving dividend payments from its offshore subsidiaries. This decision marked one of the initial steps in what would become known as the U.S. government's "maximum pressure" strategy approach to Venezuela, which would come to include sanctions on the country's oil, gold and banking sector and secondary sanctions on non-U.S. actors involved in dealings with the Venezuelan state.[15]

As illustrated in Figure 1, during this period, Venezuela also saw the largest economic collapse documented outside of wartime in the past seventy years, with per capita income dropping by 72% - the equivalent of more than three Great Depressions.[16] The poverty rate skyrocketed to 93%, and more than 7 million Venezuelans - or nearly a fourth of the country's population – emigrated during this period.[17]

---

[15] President of the United States (2017) and U.S. Department of the Treasury (n.d., 2019a, 2019b, 2019c).
[16] Rodríguez and Imam (2022).

---

[17] R4V (2022) and ENCOVI (2022).

**Figure 1. Oil production and per capita income in selected countries.**

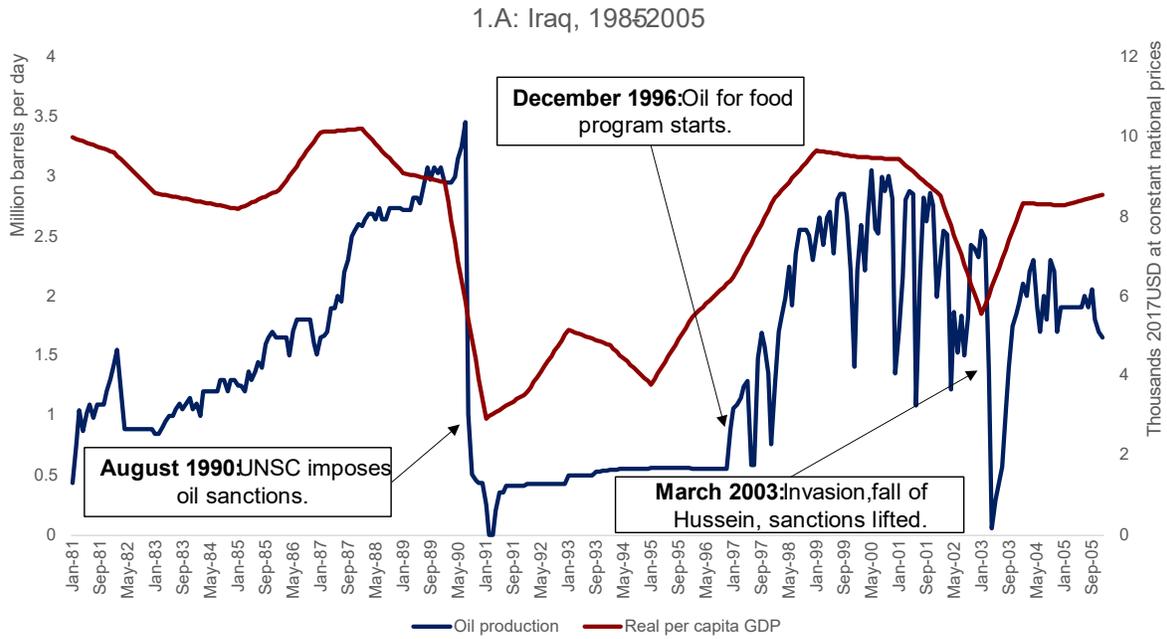

1.A: Iraq, 1981–2005

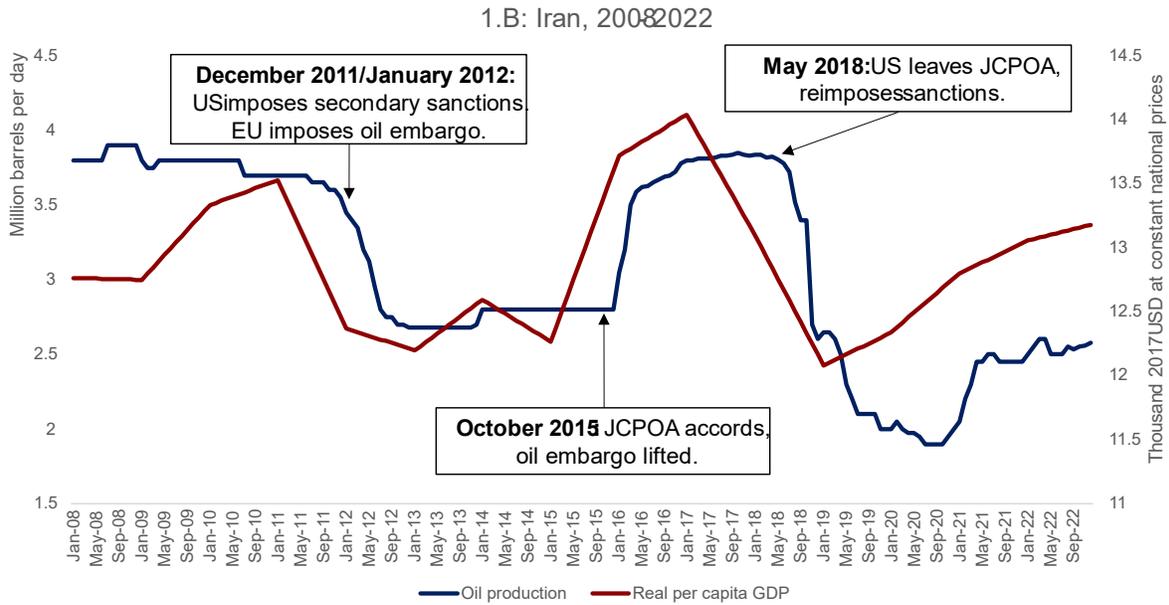

1.B: Iran, 2008–2022

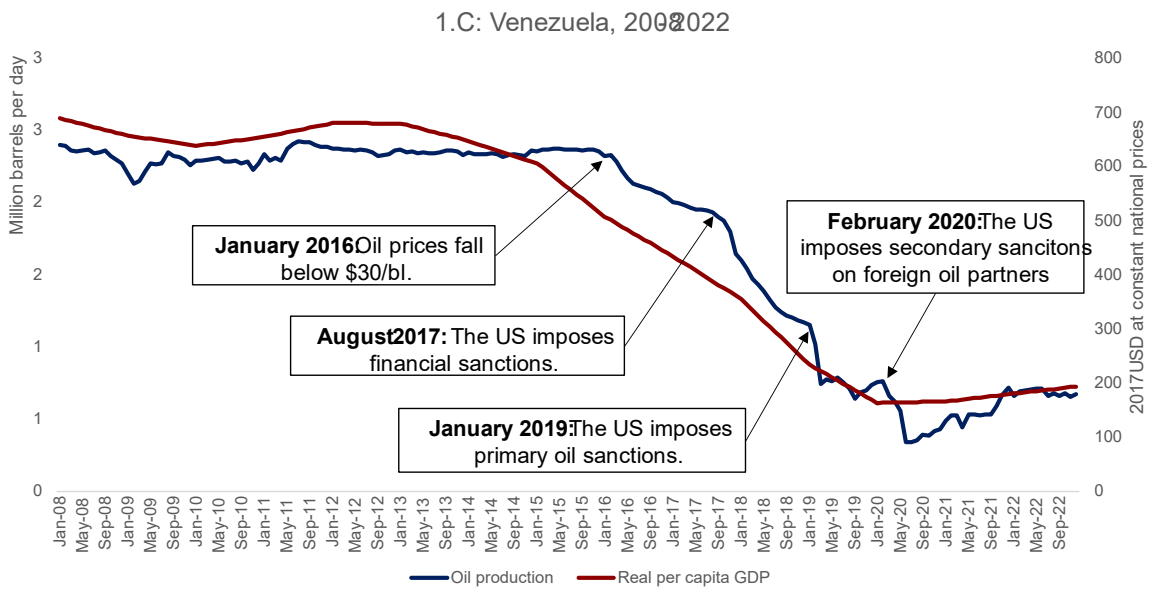

1.C: Venezuela, 2008–2022

The strong temporal coincidence between the imposition of economic sanctions and inflexion points in the country's oil production data suggests that oil sanctions played an important role in accounting for the country's economic collapse. This result is borne out in the more detailed time-series analysis investigating the effects of oil sanctions.[18]

Nevertheless, it is always possible to sketch alternative hypotheses about potential drivers of deteriorating outcomes by pointing to phenomena that occurred at roughly the same time as sanctions. For example, point to the appointment of General Manuel Quevedo as president of the state-owned oil company PDVSA in December of 2017 as evidence that changes in management could be the primary driver of the subsequent decline in the country's oil output.[19]

I address this issue by applying the quasi-experimental difference-in-differences method to study the evolution of oil production in joint ventures operating in the country's Orinoco Basin.[20] The method compares changes over time in the evolution of output in a group that was affected by sanctions (firms with prior access to financing) to a control group that was unaffected (firms without prior access). Since both groups of firms were also subject to other shocks that affected the oil industry as a whole, the method allows us to concentrate on the channel of causation of interest (financial access) while sweeping out the effect of other potential causes. The results show that firms that had access to finance in the form of loans from their foreign partners prior to the sanctions were more affected by the closing of access to external finance imposed by financial sanctions than those that lacked that access, confirming the hypothesis that severing the country's access to international financial markets significantly affected its capacity to generate export revenues.

Another argument that is commonly voiced in attempts to discount the effect of sanctions on the Venezuelan economy points to the fact that the decline in Venezuela's economic growth began before sanctions were implemented. Therefore, these critics argue, sanctions cannot be the explanation for the country's economic collapse.[21]

As we have already warned, temporal precedence arguments need to be approached carefully if we want to use them to draw conclusions about causality. One pitfall of using them in a country-level setting is that they presume that the causes of the pre-sanctions deceleration must be the same as those of the post-sanctions deceleration. In other words, using the pre-sanctions trend in output to identify the effect of sanctions implicitly assumes a counterfactual scenario in which the economy would have continued deteriorating at the same rate even if sanctions had not been imposed.

This is not a reasonable counterfactual for Venezuela because we know that one of the key drivers of the country's growth, external oil market conditions, was radically different in the pre- and post-sanctions period. There is not much doubt as to what caused Venezuela's 2014-16 recession: a 66% decline in oil prices. Plummeting oil revenues led to deep cuts in government spending and imports, fueling a decline in real wages and economic activity.

In contrast, oil prices rose by 78% between 2016 and 2018. In this period, Venezuela's economic contraction was not driven by falling oil prices but rather by falling oil production – which, as we have seen, was strongly impacted by sanctions. Had Venezuelan oil export volumes remained unchanged, this increase in the terms of trade would have generated a boost in revenues equal to nearly one-third of GDP. Under normal conditions, such a positive terms of trade shock would have guaranteed a substantial economic recovery.

---

[18] Rodríguez (2019, 2022b), Oliveros (2020 and Guerrero (2022).
[19] Bahar et. al. (2019).
[20] Rodríguez (2022b).
[21] Bahar et al. (2019).

Given what we know about the evolution of oil prices and the sensitivity of Venezuelan growth to oil market conditions, the reasonable counterfactual for the post-2017 period in the absence of sanctions is not one of a continuing deterioration but rather one of economic recovery or, at the very least, of economic stabilization. That Venezuela suffered a massive economic contraction in a period of rising oil prices suggests that it is necessary to bring in other factors, such as the imposition of economic sanctions, to explain why the historically strong correlation between oil prices and domestic economic conditions broke down precisely around late 2017.

### 6. The way ahead

Country case studies can play an important role in helping us map the effects that the severing of links to the global economy has on living standards in target countries. In-depth analysis of country experiences can lead to a clearer understanding of how the different channels of causation through which economy-wide intervention can affect living standards operate in practice, often shedding light on phenomena that the cross-national data is not well equipped to assess. If cross-country empirics are like the lamppost that throws light only on part of a sidewalk, case studies can act as the flashlight that allows us to look in more detail in areas in which a wider look is simply not feasible.

For these case studies to help further our understanding of how sanctions work we must see them as complements rather than substitutes to the cross-national evidence. One way to do so is by ensuring that case studies focus on mapping the causal channels through which economic sanctions impact an economy's living standards. What role, for example, do reductions in the government's capacity to pay for public goods and services play in the deterioration of living standards? What are the sectors most impacted by sanctions? How do we separate the effect of sanctions from those of the lack of willingness of international actors to do business with countries that are perceived as higher-risk?

The cross-national evidence can also serve as a useful check on the plausibility of alternative hypotheses to explain a country's deterioration in living standards. For example, poor policies are often put forward as an alternative explanation to account for the deep recessions suffered by many countries impacted by sanctions. One first question to ask is whether the depth of the recession seen in the sanctioned country is comparable to those observed in other countries with similarly poor policies. Alternatively, we can use the coefficient estimates drawn from cross-national empirical growth literature to assess whether the magnitude of a country's growth deterioration is consistent with the observed changes in policies.

Both experimental and quasi-experimental methods may also be useful in understanding how particular channels of causation work in a country-level setting. One promising area for future research would be to use experimental research designs similar to those in the literature on race and gender discrimination to assess the extent to which the access of firms or persons to financial services is affected by whether their country of origin is the target of sanctions.[22]

---

[22] List and Rasul (2011) and Kübler et al. (2018).